\documentclass[useAMS,usenatbib]{mn2e}
\usepackage{graphicx,amssymb}

\usepackage{times}

\newcommand{\sm}[1]{`#1'} 

\title[Massloss of galaxies due to a UV-background]{Massloss of galaxies due to a UV-background}
\author[T. Okamoto, L. Gao and T. Theuns]{Takashi Okamoto$^{1}$\thanks{E-mail:
takashi.okamoto@durham.ac.uk}, Liang Gao$^{1, 2}$ and Tom Theuns$^{1, 3}$\\
$^{1}$Institute for Computational Cosmology, Department of Physics,
Durham University, South Road, Durham, DH1 3LE\\
$^{2}$National Astronomical Observatories, Chinese Academy of Science,
Beijing, 100012, China\\
$^{3}$Department of Physics, University of Antwerp, Campus
Groenenborger, Groenenborgerlaan 171, B-2020 Antwerp, Belgium}
\begin{document}
\date{}
\pagerange{\pageref{firstpage}--\pageref{lastpage}} \pubyear{2008}

\maketitle

\label{firstpage}

\begin{abstract}
We perform cosmological hydrodynamic simulations to determine to what extent galaxies lose their gas due to photoheating from an ionizing background. We find that the characteristic mass at which haloes on average have lost half of their baryons is
$M_{\rm c}\sim 6.5 \times 10^9 \,h^{-1} \ {\rm M_\odot}$ at $z = 0$, which corresponds to a circular
  velocity of $25 \ {\rm km} \ {\rm s}^{-1}$. This is significantly lower than the filtering mass obtained by the linear theory, which is often used in semianalytical models of galaxy formation. We demonstrate it is the gas temperature at the virial radius which determines whether a halo can accrete gas. A simple model that follows the merger history of the dark matter progenitors, and where gas accretion is not allowed when this temperature is higher than the virial temperature of the halo, reproduces the results from the simulation remarkably well. This model can be applied to any reionization history, and is easy to incorporate in semianalytical models. 
\end{abstract}


\section{Introduction}

A UV-background produced by quasars and stars acts to quench
star formation in small galaxies by photoheating their gas, which gets too hot to be confined in their shallow potential wells. This was originally pointed out by \citet{dzn67}, and 
investigated in the context of the cold dark matter (CDM) model by
\citet{cr86}. Depending on the spectral shape of the ionizing background, the typical virial temperature below 
which haloes lose their gas is around $10^4$~K. Photoheating associated with reionization therefore inhibits star formation in
dwarf galaxies and affects the faint end of the galaxy luminosity
function \citep[e.g.][]{efs92}.\\

Since \cite{rees86} argued that in dark matter haloes with circular
velocities around 30 km s$^{-1}$ gas can be confined in a stable
fashion, with radiative cooling balancing photoheating, many attempts
to quantify the effects of an ionizing background on galaxy formation
have been made using semianalytical calculations \citep[e.g.][]{br92,
  efs92, sgb94, ngs99, bul00, ben02a, ben02b, som02}, spherically
symmetric simulations \citep{tw96, kit00}, and three-dimensional
cosmological hydrodynamic simulations \citep{qke96, ns97, whk97,
  gne00}. The effects of self-shielding have been investigated using
three-dimensional radiative hydrodynamic simulations by \citet{su04a,
  su04b}.\\

However there is still debate about the value of the characteristic
mass, $M_{\rm c}$, below which galaxies are strongly affected by
photoionization. \citet{gne00} argued that $M_{\rm c}=M_{\rm F}$, the {\it
  filtering mass} that corresponds to the scale over which baryonic
perturbations are smoothed in linear perturbation theory (see Appendix~\ref{FILTERINGMASS}).
\citet{hoe06} used simulations to argue that $M_{\rm c}\ll M_{\rm F}$, in
particular at low redshift. They argued that $M_{\rm c}$ follows from
considering the equilibrium temperature, $T_{\rm eq}$, between
photoheating and radiative cooling at a characteristic overdensity of
$\Delta \simeq 1000$; $M_{\rm c}$ haloes have a virial temperature $T_{\rm
  vir} \simeq T_{\rm eq}(\Delta = 1000)$.\\

The relation $M_{\rm c}=M_{\rm F}$ proposed by \citeauthor{gne00} is often used in
semianalytic modelling for dwarf galaxy formation in order to describe
the quenching of star formation \citep{bul00, ben02a, ben02b, som02}. An important
application of this relation is to estimate whether reionization can
explain the apparent dearth of satellite galaxies in the Milky Way, as
compared with the high abundance of dark matter subhaloes \citep{moo99}.
Interestingly, \citet{som02} and \citet{no06} argued that models underestimate 
the number of dwarf satellite galaxies in the Local Group if the effect is 
as strong as inferred from the filtering mass. 
\\

Given the uncertainty in $M_{\rm c}$, we were motivated to perform high-resolution hydrodynamic simulations of galaxy
formation in a $\Lambda$CDM universe with a time-evolving
UV-background, and to measure the evolution of the baryon fraction as
function of halo mass. We compare our results to those of \citeauthor{gne00}  and 
\citeauthor{hoe06}, and formulate an intuitive and simple model
that reproduces our simulations very well. The model can be
incorporated easily into any semianalytic model of galaxy formation.\\

This paper is organised as follows. Details of our simulations are
described in Section~2. Section~3 contains our numerical results which
focus on the baryon fraction of the simulated haloes, and the
characteristic mass at which the baryon fraction is on average half the
cosmic mean. We present a simple physical model that describes the
evolution of the baryon fraction of individual haloes and compare it to
the simulations in Section~4. Discussion and a summary follow in
Sections~5 and 6.

\section{Simulations}

\subsection{Background cosmology and radiative processes}

The physics of how low-mass haloes fail to accrete gas, or lose it
after the gas is being ionized, is largely independent of cosmology, but for
definiteness we assume a geometrically flat, low density CDM universe
($\Lambda$CDM) with cosmological parameters: $\Omega_0 = 0.25$,
$\Omega_{\Lambda} \equiv \Lambda_0/(3 H_0^2) = 0.75$, $\Omega_{\rm b} =
0.045$, $h \equiv H_0/(100$ km s$^{-1}$ Mpc$^{-1}) = 0.73$, and
$\sigma_8 = 0.9$.  Symbols have their usual meaning.\\

The radiative processes that we take into account are Compton cooling
off the microwave background, thermal Bremsstrahlung cooling, and line
cooling and photoionization heating from Hydrogen and Helium in the
presence of an imposed ionizing background as computed by \citet{hm01}. 
Although our cooling routine handles cooling and
photoheating for many species, we assume primordial gas throughout this
paper for simplicity. The routine was developed for a different project
and will be described elsewhere\footnote{We would like to thank our
  colleagues J. Schaye, C. Dalla Vecchia and R. Wiersma for allowing us to
  use these rates.}. Briefly it employs tabulated reaction rates
evaluated with {\small CLOUDY} \citep{cloudy}, assuming ionization
equilibrium. We assume Hydrogen and Helium I reionization occurs at $z = 9$,
and HeII reionization at $z=3.5$.\\

We increase the thermal energy by 2 eV per atom both during H
and HeII reionization to mimic radiative transfer and non-equilibrium effects \citep{ah99}. 
This brings the evolution of $T_0$ and $\gamma_{\rm EOS}$ that characterise
the low-density temperature-density relation,
$T(\rho)=T_0\,(\rho/\langle\rho\rangle)^{\gamma_{\rm EOS}-1}$
 closer to the measurements of \citet{sch99}
\footnote{Throughout this paper, $\rho$ and $\langle \rho \rangle$ denote physical gas density 
and the mean baryon density in the universe, respectively, unless otherwise stated.}. 
The extra energy is distributed over redshifts with a half-Gaussian distribution with a
dispersion $\sigma_z = 0.0001$ from $z = 9$ for Hydrogen reionization
and with a Gaussian distribution centred on $z = 3.5$ and with
$\sigma_z = 0.5$ for HeII reionization. While our numerical results
depend on the assumed ionizing background, our model, described in
Section~\ref{OURMODEL}, can be applied to any assumed thermal history.\\

\subsection{Runs}

We use a modified version of the {\small PM-TreeSPH} code {\small
  GADGET2} \citep{gadget2}, the successor of the {\small TreeSPH} code
{\small GADGET} \citep{gadget1}. Hydrodynamics is treated with Smoothed
Particle Hydrodynamics \citep[SPH; ][]{luc77, gm77}, employing a
\sm{conservative entropy} formulation that manifestly conserves energy
and entropy \citep{sh02}.\\

\begin{table*}
\caption{Numerical parameters of the performed runs. Simulations use periodic boxes with the indicated box size, 
\lq zoomed\rq~ simulations evolve part of the simulation box at higher resolution. $m_{\rm SPH}$ and $m_{\rm DM}$ 
  denote the particle masses of SPH and 
  (high-resolution) dark matter particles, 
  respectively. $\epsilon$ is the comoving gravitational softening length in
  terms of the equivalent Plummer softening. The gravitational force
  obeys the exact $r^{-2}$ law at $r > 2.8 \epsilon$.}
\begin{center}
\begin{tabular}{@{}lcccccc}
\hline
Simulation & Box Size & Zoom & $N_{\rm SPH}$ & $m_{\rm SPH}$ & $m_{\rm DM}$ & $\epsilon$  \\
name    & ($h^{-1}$ Mpc) & & & ($10^4 h^{-1} {\rm M}_{\odot}$) & ($10^4 h^{-1} {\rm M}_{\odot}$) & ($h^{-1}$ kpc) \\
\hline
reference & 4 & No & $256^3$ & 4.76 & 21.7 & 0.5 \\
high-resolution  & 4 & Yes & $256^3$ & 0.596 & 2.71 & 0.25 \\
low-resolution  & 8 & No & $256^3$ & 38.1 & 174 & 1 \\
reference-lbox & 8 & Yes & $256^3$ & 4.76 & 21.7 & 0.5 \\
\hline
\end{tabular}
\end{center}
\label{SIMS}
\end{table*}

We have performed a series of runs which differ in numerical resolution
and box size (Table~\ref{SIMS}). We also use a run with a different
imposed ionizing background. The \sm{reference} simulation uses a
periodic cubic box of comoving size $L = 4 \ h^{-1}$~Mpc, and equal numbers of
$N=256^3$ dark matter and SPH particles.  The \sm{high-resolution} and
\sm{low-resolution} simulations are used to gauge numerical
convergence.  The \sm{high-resolution} simulation used zoomed initial
conditions, where a region of size $L_{\rm h} = 2 \ h^{-1}$ Mpc inside
the \sm{reference} box is simulated with eight times better mass
resolution. The region outside the high-resolution region is populated
with dark matter particles only, at coarse resolution.  Linear waves
common to both simulations have identical phase and amplitude. The \sm{low-resolution}
simulation uses a $8 \ h^{-1}$ Mpc comoving cube, with uniform mass
resolution, eight times coarser than the \sm{reference} simulation.
Simulation \sm{reference-lbox} has identical resolution to
\sm{reference}, using zoomed initial conditions in a larger box.\\

We convert SPH particles into collisionless (\lq star\rq) particles in
high density regions (those with gas overdensity $\Delta \equiv \rho/\langle\rho\rangle > 1000$ 
and physical Hydrogen number density $n_{\rm
  H} > 5$ cm$^{-3}$) in a time given by the local dynamical time,
$t_{\rm dyn} = (4 \pi G \rho)^{-1/2}$. This decreases the
computational cost of the simulations significantly\footnote{We do not
  include stellar feedback, since we concentrate here on radiative
  feedback only.}, and we have verified it does not change the results
as compared to simulations without star formation.\\

\begin{figure*}
\begin{center}
\includegraphics[width=15cm]{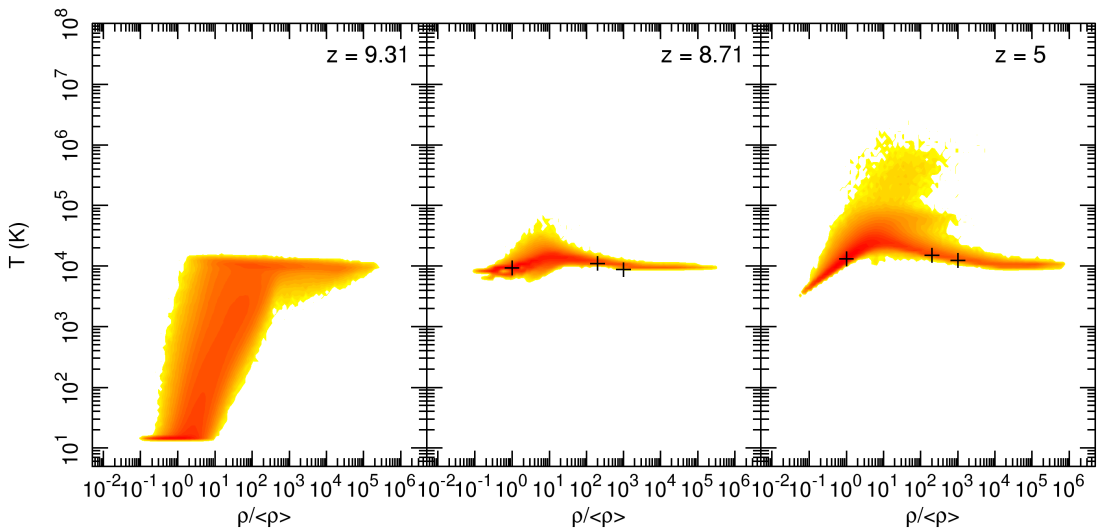}
\end{center}
\caption
{
   Distribution of the particles in the density-temperature phase
  diagram for the \sm{reference} simulation, colour coded according to the fraction of particles at given $\rho$ and $T$, at redshifts $z=9.31$ (before reionization), $z=8.71$ (just after reionization) and $z=5$, as indicated in the panels. Adiabatic cooling, impulsive heating during reionization, photoheating, radiative cooling and shocks from structure formation determine the distribution and its evolution. Plus signs indicate temperatures obtained assuming gas evolves at constant overdensity. This reproduces $T(z)$ well for the indicated overdensities of  $\Delta = 1, \ 200,$ and $1000$.
}
\label{PHASE}
\end{figure*}

The evolution of gas in the temperature-density plane is shown in
Fig.~\ref{PHASE}.  Before reionization, low density gas lies
along the adiabat $T\propto\rho^{\gamma-1}$, with $\gamma=5/3$, except
where shocked, whereas cooling introduces a sharp maximum temperature
of $T\sim 10^4$~K at higher density. Bigger simulation boxes will
contain more massive haloes which do contain shocked gas above
$10^4$~K. Heat input during reionization makes the gas nearly
isothermal at $T\sim 10^4$~K. At later times gas is above a minimum
temperature $T_{\rm min}(\rho, z)$, set by the balance between
photoheating and adiabatic cooling at low density, and photoheating
and radiative cooling at higher density \citep{hg97, the98}. 
Shocked gas forms the plume above $T_{\rm min}(\rho,z=5)$
seen in Fig.~\ref{PHASE}.\\

The equilibrium temperature $T_{\rm eq}(\rho,z)$ where photoheating
balances radiative cooling is indicated by crosses for overdensities
$\Delta = 200$ and $1000$, whereas we also include adiabatic cooling
for the cross at $\Delta=1$. As expected these points fall on the
$T_{\rm min}(\rho,z)$ curves both for $z=8.71$ (just after
reionization) and the later redshift $z=5$.\\

\section{Analysis of simulations}
\subsection{Halo identification}\label{FOF}

We identify haloes and the baryons associated with them as
follows. First we identify overdense regions using the
friends-of-friends (FoF) algorithm \citep{dav85} on the dark matter
particles only, choosing the linking according to \citet{on03}.  
This picks-out overdensities according to the spherical top-hat model
appropriate for the assumed cosmological parameters  \citep{ecf96}.\\

We refine the location of the centre of each given halo, by determining
the centre of mass of all particles within a sphere of given radius
$R$, and progressively shrinking $R$ \citep{lc94}. The location of the
density peak is identified as the centre of the halo. Finally we
determine the virial radius of the halo, $R_{\rm vir}$, such that the
mean dark matter density within $R_{\rm vir}$ equals the virial density
of dark matter component.\\

Finally, SPH and star particles within $R_{\rm vir}$ from the halo centre are assigned to that halo. 
We restrict the analysis below to haloes with at least 100 dark matter particles, and discard haloes 
contaminated by boundary particles in the zoomed simulations.\\

\subsection{Baryon fraction of simulated haloes}
\begin{figure*}
\begin{center}
\includegraphics[width=15cm]{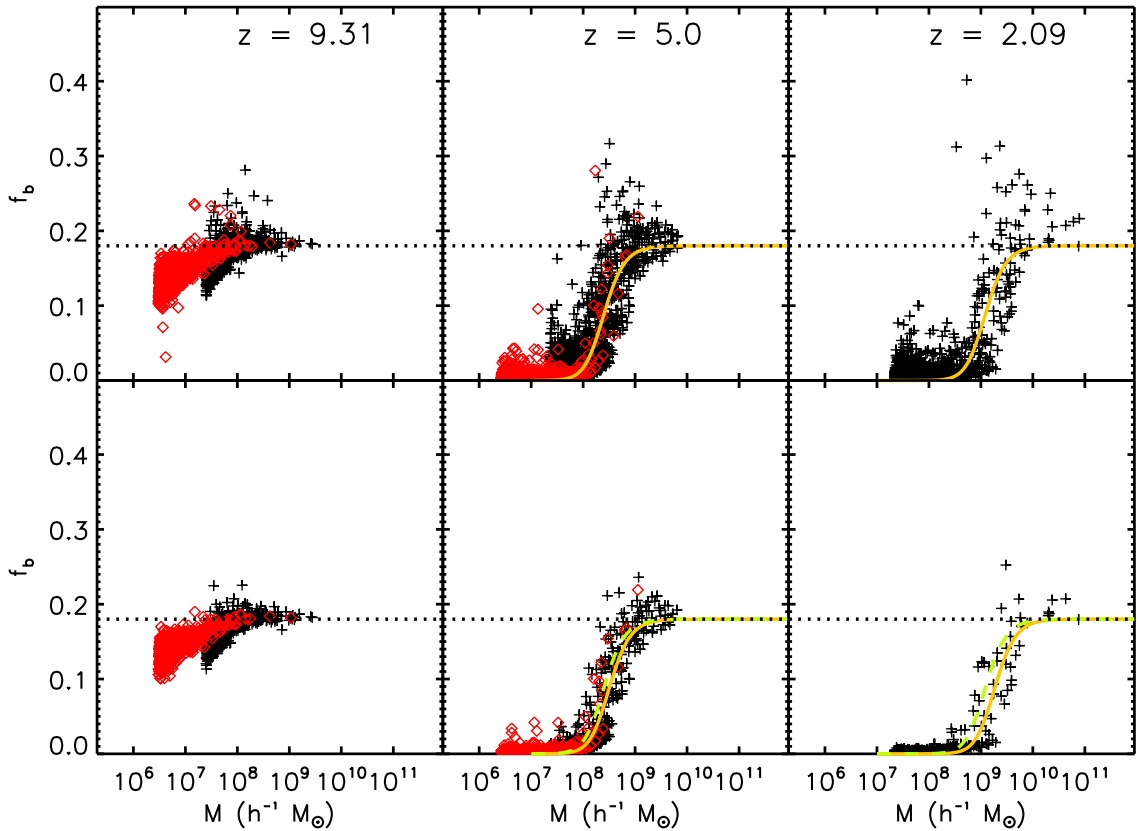}
\end{center}
\caption
{
Fraction of baryonic mass as function of total mass, $M$, for simulated haloes at three redshifts $z$ for all haloes ({\em top row}) and \lq isolated haloes\rq~ ({\em bottom row}). 
The cosmological  baryon fraction $\Omega_b/\Omega_0\approx 0.18$ is indicated as the dotted horizontal line.
Plus signs and diamonds indicate haloes from the \sm{reference} and \sm{high-resolution} simulations,
  respectively. Their comparison shows that the drop in $M_b/M$ below $M\sim 10^8M_\odot$ before reionization ($z=9$ in these simulations) is an artifact of numerical resolution; at higher masses or after reionization both simulations give similar values, demonstrating numerical convergence.
The baryon fraction $M_b/M$ falls from the cosmic mean to just a few per cent over about a decade in mass below a critical value $M_{\rm c}$ which depends on $z$. Halo to halo scatter is much reduced when considering only isolated haloes (compare top with bottom panels), which removes tidally stripped haloes.  Full lines indicate fits using equation~(\ref{FITTING}) to the \sm{reference} model; these fit the simulated haloes very well. The dashed lines in the bottom panel repeat the fits to the top row; the critical mass in isolated haloes is slightly higher than for all haloes.}
\label{BFRAC}
\end{figure*}

\begin{figure}
\begin{center}
\includegraphics[width=7.5cm]{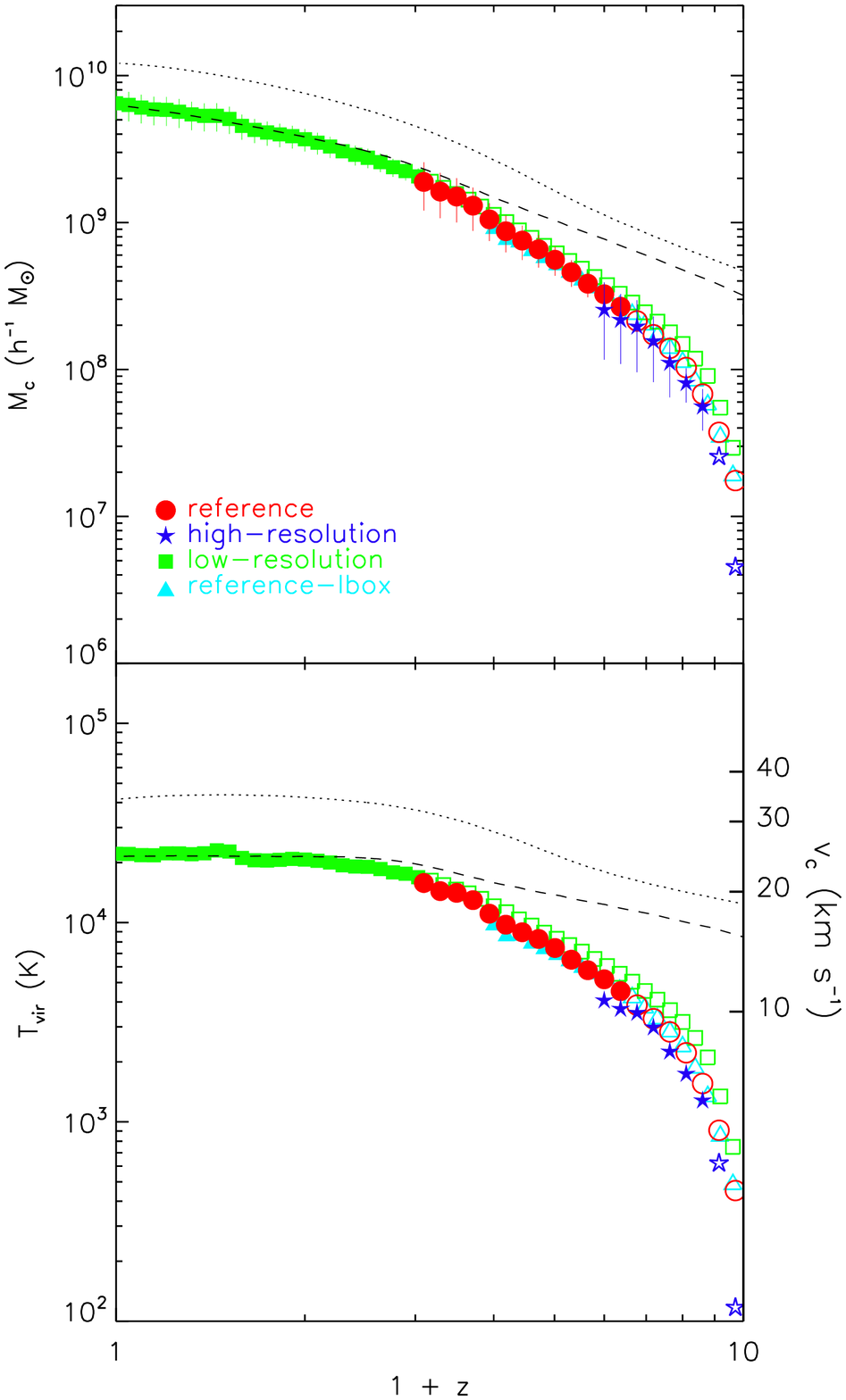}
\end{center}
\caption
{
  {\it Top panel}: Evolution of the characteristic mass $M_{\rm c}$ with redshift $z$.
  Different symbols refer to simulations with the same assumed ionizing background 
  but different resolution and/or size of the simulation box 
  ( \sm{reference}: circles; \sm{high-resolution}: stars;  \sm{low-resolution}: squares; \sm{reference-lbox}: triangles). 
  When haloes of $M=M_{\rm c}$ contain fewer than $10^3$ dark matter particles the simulation 
  is shown with open symbols. 
  Overlap of the different filled symbols demonstrates numerical convergence. 
  $M_{\rm c}$ rises from $\sim10^7\,h^{-1}M_\odot$ just after reionization to 
  $\sim 6.49\times 10^9\,h^{-1}M_\odot$ at $z=0$.  
  Lines indicate the virial masses corresponding to virial temperatures of $T_{\rm vir}=T_{\rm eq}(\Delta = 1/3\,\Delta_{\rm vir})$ 
  (dotted) and $T_{\rm eq}(\Delta = 1000)$ (dashed); the redshift dependence of these lines is quite different from that of $M_{\rm c}$, demonstrating a single temperature cut is not sufficient to compute the characteristic mass. It is necessary to follow the merger history of the haloes.
  {\it Lower panel}: corresponding evolution of the virial temperature ({\em left hand scale}) 
  and circular velocity ({\em right hand scale}). 
  $M_{\rm c}$ has been converted to $T_{\rm vir}$ using Eq.~(\ref{eq:TVIR}). 
  The characteristic circular velocity $V_c(M_{\rm c})$ rises rapidly to $\approx 10$~km s$^{-1}$ 
  following reionization, then continues to increase more gradually reaching $\approx 20$~km s$^{-1}$ at $z=2$, 
  and remains roughly constant thereafter. 
  The equilibrium temperatures at two overdensities ($\Delta = 1/3 \, \Delta_{\rm vir}$: dotted; $\Delta = 1000$: dashed) 
are also shown; they are nearly constant, whereas the virial temperature of haloes with mass $M_{\rm c}$ increases rapidly especially at early times.
}
\label{MCEVO}
\end{figure}

Before reionization the baryon fraction $f_{\rm b}\equiv M_{\rm b}/M$ of
simulated haloes scatters around the cosmic mean, $\langle f_{\rm b}\rangle\equiv \Omega_{\rm b}/\Omega_0$, as expected
(Fig~\ref{BFRAC}; the drop toward lower masses is a resolution
effect). After reionization $f_{\rm b}$ drops sharply below the
characteristic mass $M_{\rm c}(z)$; haloes with $M\le M_{\rm c}(z)$ have too
shallow potential wells to hold on to their photoheated gas. The drop of $f_{\rm b}$
from the cosmic mean to just of few per cent occurs over approximately
a decade in total mass. The characteristic mass $M_{\rm c}$ where
$f_{\rm b}=\langle f_{\rm b}\rangle/2$ is the same for \sm{reference} and
\sm{high-resolution} simulations demonstrating
numerical convergence\footnote{We stop the \sm{high-resolution} simulation at $z = 5$ because of its
small volume: the number of haloes with $M>M_{\rm c}$ becomes too small to
determine $M_{\rm c}$. We stop the \sm{reference} simulation at $z = 2.09$
for the same reason.}.\\

Some haloes have $f_{\rm b}\gg \langle f_{\rm b}\rangle$.  By tracking their
merger histories we find that most of these were subhaloes ({\em i.e.}
part of a larger halo) at earlier times, which have been tidally
stripped from their hosts leaving baryon-rich cores.  \citet{lud08}
found that many apparently isolated low-mass haloes are subhaloes
ejected by multiple-body interactions. This phenomenon of tidal
stripping may have occurred for haloes with $f_{\rm b}\le\langle f_{\rm b}\rangle$
as well; therefore the baryon fraction of such low-mass haloes may be
artificially high.\\

To reduce the contamination from tidally stripped haloes we define {\it
  isolated} haloes as those that lie outside $6 \ R_{\rm vir}$ of more
massive haloes. The baryon fraction of isolated haloes shows much less
scatter as function of halo mass (Fig.~\ref{BFRAC}, bottom panels), and we
restrict our analysis to isolated haloes\footnote{This decreases the number of
haloes from 3618 to 1185 for the \sm{reference} simulation at
$z=2.09$.}.\\

We use the following function proposed by \citet{gne00} to describe how the baryon fraction depends on mass and redshift:
\begin{equation}
f_{\rm b}(M,z) = \langle f_{\rm b}\rangle \left\{ 1 + (2^{\alpha/3}
  - 1) \left(\frac{M}{M_{\rm c}(z)}\right)^{-\alpha}
\right\}^{-\frac{3}{\alpha}}\,
\label{FITTING}
\end{equation}
in terms of the cosmic mean baryon fraction, $\langle f_{\rm b}\rangle$.
This function has two fitting parameters: $\alpha$ and $M_{\rm c}(z)$.
The baryon fraction of haloes with $M\gg M_{\rm c}$ equals the cosmic mean while that of haloes with $M\ll M_{\rm c}$ goes to zero $\propto (M/M_{\rm c})^3$.  The parameter $\alpha$ controls how rapidly $f_{\rm b}$ drops for low mass haloes;
a value of $\alpha=2$ fits the simulations well
(Fig.~\ref{BFRAC}), and is also consistent with the
results of \citet{hoe06}. Haloes with mass equal to the characteristic mass, $M=M_{\rm c}$ have lost half their baryons. As expected, the value of $M_{\rm c}$ for isolated haloes is slightly higher than for the full sample of haloes (by 27 per cent at $z = 5$ and 50 per cent at $z = 2.09$ for the \sm{reference} simulation; see Fig.~~\ref{BFRAC}).\\

The characteristic mass increases from $M_{\rm c}(z) \approx 10^7~h^{-1}M_\odot$ just after reionization to 
$M_{\rm c}\approx 6.49\times 10^{9}~h^{-1} M_\odot$ at $z=0$ (Fig.~\ref{MCEVO}). 
The  \sm{high-resolution} simulation has a smaller value of  $M_{\rm c}$ by 20 per cent as compared to the 
\sm{reference} simulation, at $z=5$. 
This may simply reflect the small number of massive haloes in the \sm{high-resolution} simulation 
(see the lower middle panel of Fig.~\ref{BFRAC}). 
Haloes with fewer than $10^3$ dark matter particles may suffer from two-body heating due to massive dark particles \citep{sw97}; 
they are shown as open symbols. 
Values of $M_{\rm c}$ for these less well-resolved haloes are only slightly higher than of haloes resolved with more particles. 
Overall though, simulations that differ in numerical resolution and/or box size give similar results, 
demonstrating numerical convergence. 
This gives confidence in the value of $M_{\rm c}(z=0)\approx 6.48\times 10^9~h^{-1}M_\odot$ 
from the \sm{low-resolution} simulation. The amazing agreement with the value of  $6.5 \times 10^9~h^{-1}M_{\odot}$ 
obtained by \citet{hoe06} for simulations of voids may be slightly fortuitous.\\

We convert halo mass $M$ to virial temperature $T_{\rm vir}$ using 
\begin{equation}
T_{\rm vir} =\frac{1}{2} \frac{\mu m_{\rm p}}{k_{\rm B}} V_{\rm c}^2,
\end{equation}
where
\begin{equation}
V_{\rm c} = \left( \frac{G M}{R_{\rm vir}} \right)^\frac{1}{2}
\end{equation}
is the circular velocity of the halo at the virial radius. While we compute the mean molecular weight $\mu$ self-consistently in our simulations,  here we use $\mu = 0.59$ to compute $T_{\rm vir}$ for simplicity. In terms of the overdensity $\Delta_{\rm vir}$ within the virial radius $R_{\rm vir}$, $T_{\rm vir}$ depends on halo mass and redshift as 
\begin{equation}
T_{\rm vir} (M,z)= \frac{1}{2} \frac{\mu m_{\rm p}}{k_{\rm B}}
\left(\frac{\Delta_{\rm vir} \Omega_0}{2}\right)^\frac{1}{3}
  (1 + z) (G\, M\, H_0)^\frac{2}{3}.
\label{eq:TVIR}
\end{equation}
The value of $T_{\rm vir}$ at the critical mass $M_{\rm c}(z)$ rises rapidly to $\approx 5000$~K following reionization, keeps increasing to $\sim 10^4$~K at $z=2$, then increases much more slowly to $\simeq 2.2 \times 10^4$ K ($V_c\simeq 25 \ {\rm km} \  {\rm s}^{-1}$) at $z=0$ (Fig.~\ref{MCEVO}, bottom panel). Our $z=0$ value is much lower than the temperature below which galaxy formation is assumed to be strongly suppressed in semianalytical models of galaxy formation \citep[e.g. $\sim 10^5 \ {\rm K}$ in][]{ben02b}.

\section{Modelling the suppression of the baryon fraction by the
  photoionizing background} \label{OURMODEL}

\begin{figure*}
\begin{center}
\includegraphics[width=15cm]{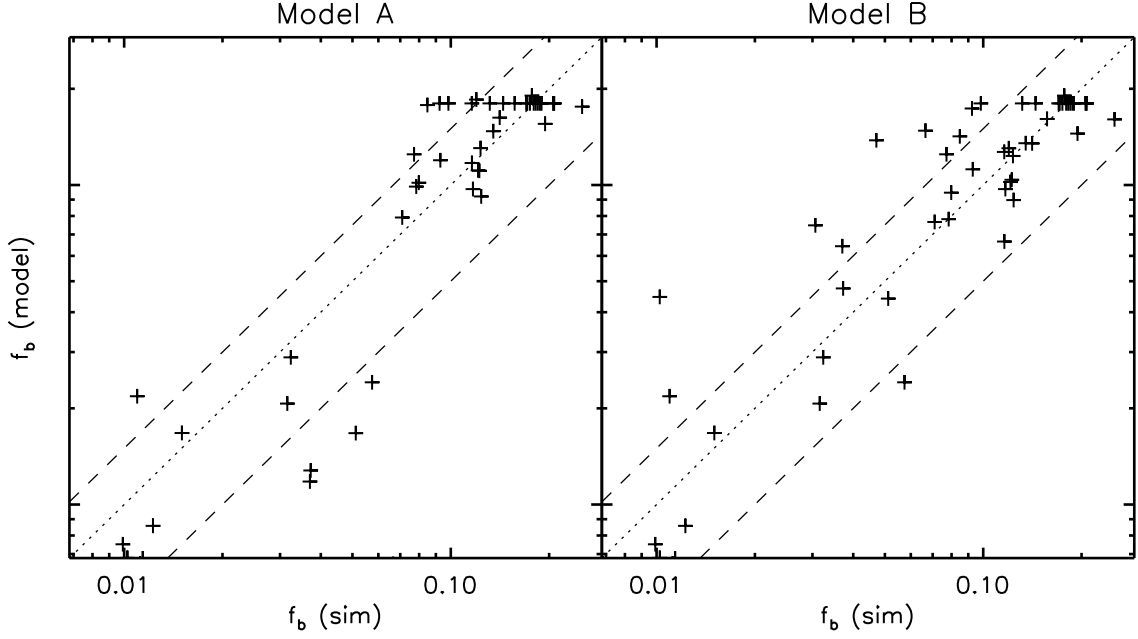}
\end{center}
\caption
{
  Baryon fraction of isolated haloes calculated by our models compared
  to values from the SPH simulation for isolated  haloes with more than 100 dark matter particles 
  and with $f_{\rm b} > f_{\rm b}^{\rm min}$ picked
  from the \sm{reference} simulation at $z = 2.09$.
  The simple model A (left hand panel) typically reproduces the baryon fraction obtained from the simulation within 50 per cent,  
  whereas model B (right hand panel) improves the agreement especially for haloes having low baryon fraction, $f_{\rm b} < 
  \langle f_{\rm b} \rangle/2$. 
  The dashed lines indicate $|f_{\rm b,\, model} - f_{\rm b, \, sim}|/ f_{\rm b, \, sim} = 0.5$. 
}
\label{HALOBYHALO}
\end{figure*}

A simple model of gas accretion compares the halo's virial temperature $T_{\rm vir}$ with the gas temperature $T$: 
if $T > T_{\rm vir}$, there is no accretion.  However the temperature $T$ of the IGM depends on its density $\rho$ (Fig.~\ref{PHASE}): 
  at which density should we evaluate $T$? 
Gnedin's filtering mass suggests to use a density close to the mean (see Appendix~\ref{FILTERINGMASS}) 
whereas \citet{hoe06} used $\rho=10^3\langle\rho\rangle$ appropriate for significantly overdense gas inside haloes.  
However, if one considers the density profile of a halo, gas at the edge of the halo (at $R_{\rm vir}$) has a local density of order 
one-third of the virial gas density, $\rho_{\rm vir}/3$. 
This suggests that gas will be prevented from accreting when $T_{\rm eq}(\rho_{\rm vir}/3)>T_{\rm vir}$.
This temperature, $T_{\rm eq}(\rho_{\rm vir}/3)$, is uniquely defined as a function of redshift\footnote{Note that the equilibrium temperature also depends on gas metallicity.} from cooling and heating 
functions used in the simulations.
\\

Simply identifying $M_{\rm c}$ from the condition $T_{\rm eq}(\rho_{\rm vir}/3) = T_{\rm vir}$ does not work well, 
overestimating $M_{\rm c}(z=0)$ by a factor $\sim 2$  as compared to the simulations (dotted line in Fig.~\ref{MCEVO}). In addition, the redshift dependence of $M_{\rm c}(z)$ is quite different as well.
Clearly this model is too simple, as it ignores the baryon fraction of the progenitors of a given halo, as well as the photoevaporation of gas from haloes during reionization. For the same reasons, the agreement at $z=0$ between a model by \citet{hoe06} which compared the equilibrium temperature at high density ($\Delta = 1000$) with $T_{\rm vir}$ (dashed line in Fig.~\ref{MCEVO}) and the simulations is probably a coincidence. Next we describe an improved model that takes the {\em merger history} of haloes into account, as well as modelling photoevaporation and the decrease in accretion rate if gas it too hot. \\

Before reionization we assume each halo has the cosmic baryon fraction, $\langle f_{\rm b}\rangle$.
We begin to construct merger trees of FoF haloes with more than 32 dark matter particles once the Universe is reionized. 
We model photoevaporation during reionization by computing the equilibrium temperature at large overdensity and comparing it to  the halo's virial temperature:
when $T_{\rm eq}(\Delta_{\rm evp})>T_{\rm vir}$ the baryon fraction decreases $\propto \exp(-t/t_{\rm evp})$, where the
evaporation  time-scale $t_{\rm evp} = {R_{\rm vir}/ c_{\rm s}}(\Delta_{\rm evp})$. 
The time $t$ is counted either starting from reionization, or from the previous merger (whichever is later), 
 $R_{\rm vir}$ is the virial radius and the sound speed $c_{\rm s}$ is evaluated at $T_{\rm eq}(\Delta_{\rm evp})$.
Since $T_{\rm eq}(\rho)$ is nearly independent of $\rho$ for high densities, the exact value of 
the overdensity is not important as long as $\Delta_{\rm evp} \gg \Delta_{\rm vir}$;  
for reference we use $\Delta_{\rm evp} = 10^6$.\\

When two haloes merge, we add the baryon masses of the progenitors, and we also allow the halo to \lq accrete\rq~ baryons, 
but only when the temperature of the accreting gas is lower than the virial temperature:
$T_{\rm eq}(\rho_{\rm vir}/3)\le T_{\rm vir}$.  
Haloes that form after reionization at the resolution limit of the FoF are assumed to have the minimum baryon 
fraction\footnote{We assume the baryon density in a halo is at least the mean baryon density in the universe. 
  Thus $f_{\rm b}^{\rm min} \equiv \langle \rho \rangle/ (\rho_{\rm vir}^{\rm DM} + \langle \rho \rangle) < 10^{-2}$. 
This gives a good estimate of the baryon fraction of low-mass haloes with $M \ll M_{\rm c}$}, $f_{\rm b}^{\rm min}$. 
This completes our model \sm{A}; 
it reproduces $f_{\rm b}(M)$ as measured from the simulations reasonable well (Fig.\ref{HALOBYHALO}, left hand panel), 
with a small tendency to under-predict $f_{\rm b}$ for haloes with $f_{\rm b}\sim 0.4$ in the simulations. 
The Appendix describes a more involved model \sm{B} that does better for such haloes
(Fig.\ref{HALOBYHALO}, right hand panel).\\

In detail, model A works as follows. Before reionization each halo has the maximum baryon mass given by the cosmic mean 
$\langle f_{\rm b}\rangle$:
\begin{equation}
M_{{\rm b}} = M_{{\rm b}}^{\rm max} \equiv {\langle f_{\rm b}\rangle\over 1 - \langle f_{\rm b}\rangle}\,M_{\rm DM}\,.
\end{equation}
The total mass of the halo is simply the sum of its baryonic and dark matter masses: $M=M_{\rm b}+M_{\rm DM}$.
After reionization the baryon mass of a halo that forms from the merger of several progenitors is
\begin{eqnarray}
M_{\rm b} &=& M'_{\rm b} + M_{{\rm acc}}\nonumber\\
M'_{\rm b} &=& \sum^{\rm prog} \exp\left(-\frac{\delta t}{t_{\rm evp}}\right) M_{{\rm b}}\,,
\label{eq::ACCRETION}
\end{eqnarray}
the sum of baryon masses of its progenitors, $M'_{\rm b}$, and the accreted baryons, $M_{\rm acc}$.
The decrease in baryon fraction $\exp(-\delta t/t_{\rm evp})$ is due to photoevaporation, with $\delta t$ 
the elapsed time since the previous merger. 
For a progenitor with $T_{\rm eq}(\Delta_{\rm evp}) < T_{\rm vir}$, we set $\delta t/t_{\rm evp} = 0$ 
since evaporation does not take place in such a halo. 
The mass available for accretion is given by 
\begin{equation}
M_{\rm acc} = \max(M_{\rm b}^{\rm max} - M'_{\rm b}, 0)\,.
\end{equation}
When the temperature of the accreting gas, $T_{\rm acc}\equiv T_{\rm eq}(\rho_{\rm vir}/3)$ is greater than the virial temperature of the halo, 
$T_{\rm vir}$, the halo does not accrete any gas: $M_{\rm acc}=0\,.$\\

\begin{figure}
\begin{center}
\includegraphics[width=8cm]{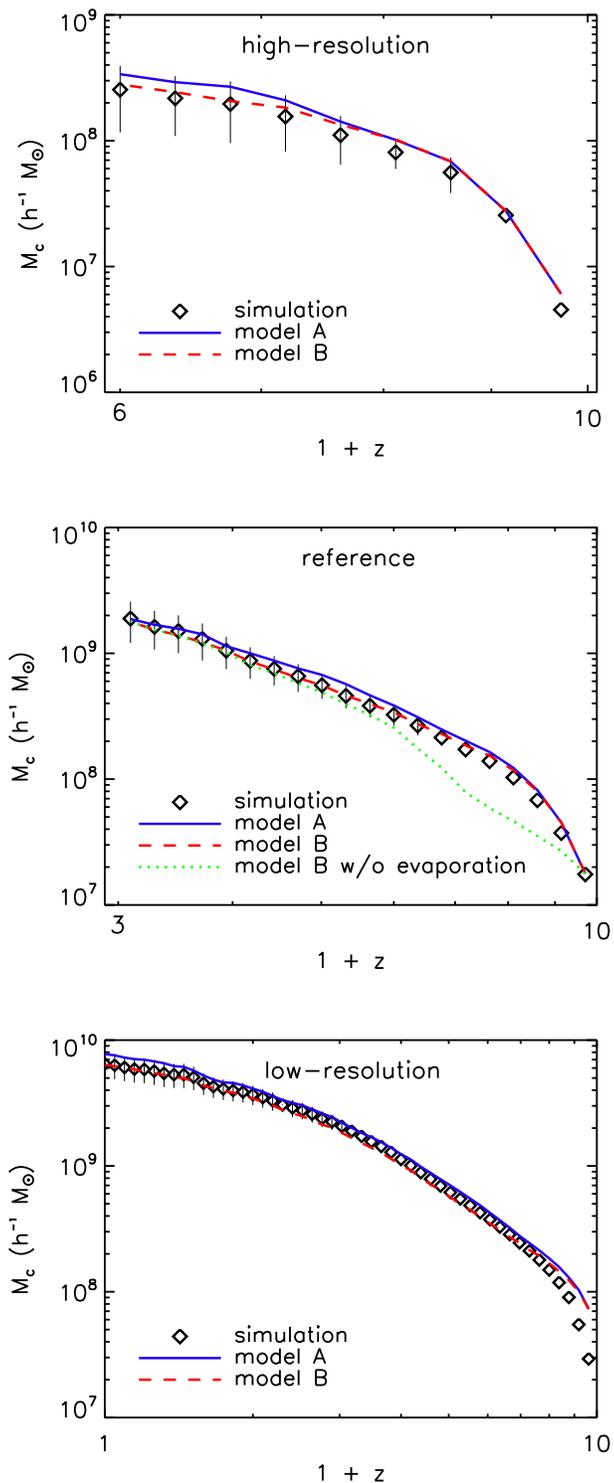}
\end{center}
\caption
{
Evolution of the characteristic mass, $M_{\rm c}$, below which haloes half lost half of their baryons, comparing numerical simulations (symbols) with the simple models A and B described in the text. These models use the dark matter merger trees with simple recipes for when haloes lose and accrete gas. Both models do well, with $M_{\rm c}(z)$ from model B  nearly indistinguishable from the full simulation. The middle panel also illustrates model B but neglecting photoevaporation during reionization: this leads to a significant underestimate in $M_{\rm c}$.\label{MODELS}}
\end{figure}

We can now predict the baryon fraction for all haloes selected from a simulation at a given redshift, and fit
Eq.~(\ref{FITTING}) to the distribution of $f_{\rm b}(M, z)$; this yields the characteristic mass $M_{\rm c}(z)$. 
This function is compared to that obtained from the simulations directly in 
Fig.~\ref{MODELS}; model A gets the simulated value to better than 20 per cent 
whereas model B is almost indistinguishable from the simulation. 
We also show model B without taking photoevaporation into account; this leads to an underestimate of $M_{\rm c}$ by more than a factor of two 
around $z\approx 6$, illustrating the importance of including this effect.\\

\section{Discussion} \label{DISCUSSION}
\begin{figure}
\begin{center}
\includegraphics[width=7.5cm]{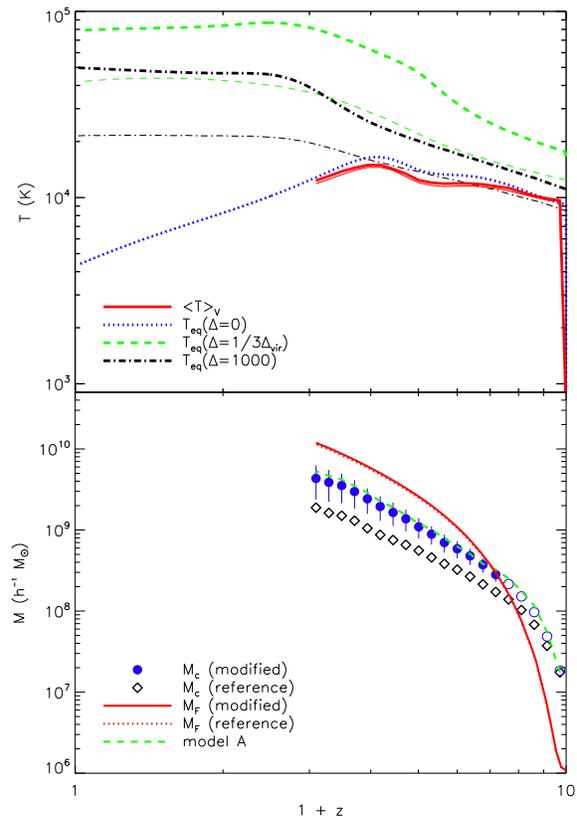}
\end{center}
\caption{ {\it Upper panel}: Temperature evolution of gas at several overdensities indicated in the legend, 
for simulations with the \citet{hm01} photoheating rate (thin lines), and an experiment with increased 
photoheating at high density (thick lines). 
{\it Lower panel}: Filtering mass, $M_{\rm F}$, computed from $\langle T \rangle_{\rm V}$ 
and characteristic mass, $M_{\rm c}$, for the original photoheating rate (dotted line and circles) 
and modified photoheating rate (full line and diamonds). 
(Open circles indicate that haloes of total mass $M_{\rm c}$ do not contain a sufficient number of dark matter 
 particles for reliable measurements.) 
 Modifying the photoheating rate does not change $M_{\rm F}$, but does increase $M_{\rm c}$ significantly. 
 Model A with the modified heating rate (dashed line) reproduces this increase well. 
 (Model B is not shown; it gives nearly identical results to model A).}
\label{MODHEAT}
\end{figure}

Our numerical results are in good agreement with those of \citet{hoe06} who used the same simulation code ({\small GADGET2})
but a different implementation of photoionization and cooling (\citet{hm96} UV-background with reionization at $z = 6$). 
The characteristic mass, $M_{\rm c}$, at which haloes lose half of their baryons is much less than the value found by \cite{gne00}: 
we find that reionization is not effective at suppressing gas accretion in haloes of the filtering mass $M_{\rm F}$ 
(Appendix~\ref{FILTERINGMASS}), only haloes with $M\le M_{\rm c}\ll M_{\rm F}$ are strongly affected. 
Our simulations suggest $M_{\rm c}$ is mainly determined by the temperature of the accreting gas at a density of one-third of 
the virial density, $\Delta \simeq 1/3\Delta_{\rm vir}$, whilst \cite{gne00} argued it is the temperature of the IGM at 
the mean density\footnote{Gnedin used the volume-averaged temperature $\langle T \rangle_{\rm V}$ instead of $T_0$}, $T_0$. 
The values of $T_0$ in his and our simulations are in fact quite similar. 
A much stronger effect found by \cite{gne00} was probably due to insufficient resolution (N. Gnedin, private communication). 

It is however important to note that radiative transfer may lead to spectral hardening and may preferentially heat 
gas in overdense regions. 
To mimic this effect, we ran a simulation where we artificially increased 
the photoheating rate by a factor of 10, for gas at overdensity $\Delta>10$:

\begin{equation}
{\cal H}_{\rm mod}(\Delta, z) =
\left\{
\begin{array}{ll}
{\cal H}(\Delta, z) &  \Delta < 10 \\
10 {\cal H}(\Delta, z)  & \Delta  \ge 10\,.
\end{array}
\right.
\label{PHOTOHEAT}
\end{equation}
${\cal H}(\Delta,z)$ is the original rate at overdensity $\Delta$ and redshift $z$, and ${\cal H}_{\rm mod}(\Delta, z)$ 
is the modified rate. 
This arbitrary modification heats gas at $\Delta = 1/3 \Delta_{\rm vir}$ to much higher temperatures yet does not 
affect $T_0$ (Fig.~\ref{MODHEAT}). 
Since $T_0$ is not affected, neither is the filtering mass. However since the accreting gas is hotter, our model 
predicts that $M_{\rm c}$ will be increased.\\

The volume-averaged temperature $\langle T \rangle_{\rm V}$ in the \sm{reference} simulation
with the modified photoheating is shown in the upper panel of
Fig.~\ref{MODHEAT} and is as expected almost identical to that in the
original \sm{reference} simulation. Consequently, the filtering
mass obtained with the modified photoheating is indistinguishable
from that in the \sm{reference} simulation even if we compute $M_{\rm F}$ from 
$\langle T \rangle_{\rm V}$ instead of $T_0$ (lower panel). 
The characteristic mass is however significantly higher, as predicted by our model.\\

This numerical experiment provides three important insights:
\begin{itemize}
\item The mean IGM temperature is {\it not important} for the evolution of the
  baryon fraction of the haloes. Hence, any agreement between the
  characteristic and filtering mass must be a coincidence. 
\item The characteristic mass is sensitive to the temperature at the
  overdensity around $\Delta = 1/3\Delta_{\rm vir}$, and therefore spectral hardening from
  radiative transfer effects can, in principle, increase the
  characteristic mass significantly. 
\item Our model works well for any reionization histories as long as
  the gas temperature as a function of density and redshift is known.
\end{itemize}

The modified heating rate used above probably significantly over estimates the effect of spectral hardening.
Although radiative transfer effects may be important during reionization, their effect will lessen after the 
gas is ionized. 
Therefore we believe our results with the original heating rate are robust, at least for lower redshifts ($z \le 6$).

  Interestingly, \citet{rt06} found a good agreement between their simulations 
    and their model on the baryon budget, which was based on the filtering mass. 
The reason for this agreement is probably that the minimum halo mass above which 
baryons are able to fall into the dark matter potential well was controlled by 
the numerical resolution rather than the filtering mass as they discussed. 
Although their highest resolution simulations well resolved the haloes whose mass 
is below the filtering mass, these simulations were stopped just after the reionization 
where the discrepancy between the filtering mass and the characteristic mass is small 
(see Appendix~\ref{FILTERINGMASS}). 
We expect their model can be greatly improved by using our model of gas accretion and evaporation.

\section{Conclusions}
We have performed high-resolution cosmological hydrodynamical simulations with
an imposed, spatially uniform but time dependent UV-background to 
investigate to what extent haloes lose baryons due to photoevaporation.
Our numerical results are consistent with the simulations of void regions by
\citet{hoe06}, but inconsistent with those found earlier by 
\cite{gne00}.  The effect of a photoionizing background on galaxy formation is much
weaker than previously assumed in semianalytic models
\citep[e.g.][]{bul00, ben02a, ben02b, som02}  based on the filtering
mass approach proposed by \citet{gne00} (see Appendix~\ref{FILTERINGMASS}).\\

We have presented a novel model which uses the merger tree of each halo, and very 
simple physical arguments about gas retention and accretion, to predict how much 
the baryonic mass is in haloes of given total mass. 
The model is in excellent agreement with the simulations. 
Gas accretion is modelled very accurately if one simply assumes that gas at the outskirts of 
the halo, at an overdensity of a third of the virial gas density, cannot accrete if it is hotter 
than the halo's virial temperature. 
A slightly more involved model does even better in predicting baryon fractions as function of 
halo mass and merger history, as compared to simulations. 
These models allow us to predict the baryon fraction of individual haloes and hence the characteristic 
mass $M_{\rm c}(z)$, at which haloes lose half of their baryons, for any assumed reionization history, 
by computing the predicted baryon fraction as a function of mass and redshift. 
The predicted $M_{\rm c}(z)$ evolution is in very good agreement with the full simulations.\\

Our simulations and models both neglect self-shielding \citep{su04a}, and radiative transfer and 
non-equilibrium effects \citep{ah99}. 
We have performed a numerical experiment in which we artificially increase the photoheating rate 
at higher density, to estimate how our results might be changed by these effects. 
The modified heating rate does not change the mean IGM temperature and hence also the filtering 
mass remains the same, yet $M_{\rm c}$ determined from the simulations increases significantly. 
Our models correctly predict this increase, demonstrating that the key physical quantity
regulating the baryon fraction in haloes is neither the mean
temperature of the IGM favoured by \cite{gne00}, nor temperature of gas
of $\Delta \simeq 1000$ favoured by \citeauthor{hoe06}, but the temperature
of the gas hanging around haloes ($\Delta \simeq 1/3 \Delta_{\rm vir}$). \\

Our models are easy to implement in any semianalytic model of galaxy formation, provide one can compute 
the equilibrium temperature at $\Delta = 1/3 \Delta_{\rm vir}$ as a function of redshift for 
the assumed UV-background.

\section*{Acknowledgments}
We are grateful to Volker Springel for providing us with the {\small
  GADGET2} code and to Joop Schaye, Claudio Dalla Vecchia, and Rob
Wiersma for providing us with tabulated radiative cooling and heating
rates. 
We thank Nick Gnedin and Masahiro Nagashima for their careful reading. 
We acknowledge support from a STFC rolling grant. The
simulations were carried out at the Cosmology Machine at the ICC,
Durham.


\begin{thebibliography}{}

\bibitem[\protect\citeauthoryear{Abel 
  \& Haehnelt}{1999}]{ah99} Abel T., Haehnelt M.~G., 1999, ApJ, 520, L13 

\bibitem[\protect\citeauthoryear{{Babul} \& {Rees}}{{Babul} \&
  {Rees}}{1992}]{br92}
{Babul} A.,  {Rees} M.~J.,  1992, MNRAS, 255, 346

\bibitem[\protect\citeauthoryear{{Benson}, {Frenk}, {Lacey}, {Baugh} \&
  {Cole}}{{Benson} et~al.}{2002a}]{ben02a}
{Benson} A.~J.,  {Frenk} C.~S.,  {Lacey} C.~G.,  {Baugh} C.~M.,    {Cole} S.,
  2002, MNRAS, 333, 177

\bibitem[\protect\citeauthoryear{{Benson}, {Lacey}, {Baugh}, {Cole} \&
  {Frenk}}{{Benson} et~al.}{2002b}]{ben02b}
{Benson} A.~J.,  {Lacey} C.~G.,  {Baugh} C.~M.,  {Cole} S.,    {Frenk} C.~S.,
  2002, MNRAS, 333, 156

\bibitem[\protect\citeauthoryear{{Bullock}, {Kravtsov} \& {Weinberg}}{{Bullock}
  et~al.}{2000}]{bul00}
{Bullock} J.~S.,  {Kravtsov} A.~V.,    {Weinberg} D.~H.,  2000, ApJ, 539, 517

\bibitem[\protect\citeauthoryear{{Couchman} \& {Rees}}{{Couchman} \&
  {Rees}}{1986}]{cr86}
{Couchman} H.~M.~P.,  {Rees} M.~J.,  1986, MNRAS, 221, 53

\bibitem[\protect\citeauthoryear{{Davis}, {Efstathiou}, {Frenk} \&
  {White}}{{Davis} et~al.}{1985}]{dav85}
{Davis} M.,  {Efstathiou} G.,  {Frenk} C.~S.,    {White} S.~D.~M.,  1985, ApJ,
  292, 371

\bibitem[\protect\citeauthoryear{{Doroshkevich}, {Zel'Dovich} \&
  {Novikov}}{{Doroshkevich} et~al.}{1967}]{dzn67}
{Doroshkevich} A.~G.,  {Zel'Dovich} Y.~B.,    {Novikov} I.~D.,  1967, Soviet
  Astronomy, 11, 233

\bibitem[\protect\citeauthoryear{{Efstathiou}}{{Efstathiou}}{1992}]{efs92}
{Efstathiou} G.,  1992, MNRAS, 256, 43P

\bibitem[\protect\citeauthoryear{{Eke}, {Cole} \& {Frenk}}{{Eke}
  et~al.}{1996}]{ecf96}
{Eke} V.~R.,  {Cole} S.,    {Frenk} C.~S.,  1996, MNRAS, 282, 263

\bibitem[\protect\citeauthoryear{{Ferland}, {Korista}, {Verner}, {Ferguson},
  {Kingdon} \& {Verner}}{{Ferland} et~al.}{1998}]{cloudy}
{Ferland} G.~J.,  {Korista} K.~T.,  {Verner} D.~A.,  {Ferguson} J.~W.,
  {Kingdon} J.~B.,    {Verner} E.~M.,  1998, PASP, 110, 761

\bibitem[\protect\citeauthoryear{{Gingold} \& {Monaghan}}{{Gingold} \&
  {Monaghan}}{1977}]{gm77}
{Gingold} R.~A.,  {Monaghan} J.~J.,  1977, MNRAS, 181, 375

\bibitem[\protect\citeauthoryear{{Gnedin}}{{Gnedin}}{2000}]{gne00}
{Gnedin} N.~Y.,  2000, ApJ, 542, 535

\bibitem[\protect\citeauthoryear{{Gnedin} \& {Hui}}{{Gnedin} \&
  {Hui}}{1998}]{gh98}
{Gnedin} N.~Y.,  {Hui} L.,  1998, MNRAS, 296, 44

\bibitem[\protect\citeauthoryear{Haardt 
  \& Madau}{1996}]{hm96} Haardt F., Madau P., 1996, ApJ, 461, 20 

\bibitem[\protect\citeauthoryear{{Haardt} \& {Madau}}{{Haardt} \&
  {Madau}}{2001}]{hm01}
{Haardt} F.,  {Madau} P.,  2001, in {Neumann} D.~M.,  {Tran} J.~T.~V.,  eds,
  Clusters of Galaxies and the High Redshift Universe Observed in X-rays
  {Modelling the UV/X-ray cosmic background with CUBA}

\bibitem[\protect\citeauthoryear{{Helly}, {Cole}, {Frenk}, {Baugh}, {Benson},
  {Lacey} \& {Pearce}}{{Helly} et~al.}{2003}]{hel03}
{Helly} J.~C.,  {Cole} S.,  {Frenk} C.~S.,  {Baugh} C.~M.,  {Benson} A.,
  {Lacey} C.,    {Pearce} F.~R.,  2003, MNRAS, 338, 913

\bibitem[\protect\citeauthoryear{{Hoeft}, {Yepes}, {Gottl{\"o}ber} \&
  {Springel}}{{Hoeft} et~al.}{2006}]{hoe06}
{Hoeft} M.,  {Yepes} G.,  {Gottl{\"o}ber} S., {Springel} V.,  2006, MNRAS,
  371, 401

\bibitem[\protect\citeauthoryear{Hui 
    \& Gnedin}{1997}]{hg97} Hui L., Gnedin N.~Y., 1997, MNRAS, 292, 27 

\bibitem[\protect\citeauthoryear{{Kitayama}, {Tajiri}, {Umemura}, {Susa} \&
  {Ikeuchi}}{{Kitayama} et~al.}{2000}]{kit00}
{Kitayama} T.,  {Tajiri} Y.,  {Umemura} M.,  {Susa} H.,    {Ikeuchi} S.,  2000,
  MNRAS, 315, L1

\bibitem[\protect\citeauthoryear{{Lacey} \& {Cole}}{{Lacey} \&
  {Cole}}{1994}]{lc94}
{Lacey} C.,  {Cole} S.,  1994, MNRAS, 271, 676

\bibitem[\protect\citeauthoryear{{Lucy}}{{Lucy}}{1977}]{luc77}
{Lucy} L.~B.,  1977, AJ, 82, 1013

\bibitem[\protect\citeauthoryear{{Ludlow}, {Navarro}, {Springel}, {Jenkins},
  {Frenk} \& {Helmi}}{{Ludlow} et~al.}{2008}]{lud08}
{Ludlow} A.~D.,  {Navarro} J.~F.,  {Springel} V.,  {Jenkins} A.,  {Frenk}
  C.~S.,    {Helmi} A.,  2008, ArXiv:801.1127

\bibitem[\protect\citeauthoryear{Moore et al.}{1999}]{moo99} 
  Moore B., Ghigna S., Governato F., Lake G., Quinn T., Stadel J., Tozzi P., 
  1999, ApJ, 524, L19 

\bibitem[\protect\citeauthoryear{{Nagashima}, {Gouda} \& {Sugiura}}{{Nagashima}
  et~al.}{1999}]{ngs99}
{Nagashima} M.,  {Gouda} N.,    {Sugiura} N.,  1999, MNRAS, 305, 449

\bibitem[\protect\citeauthoryear{Nagashima 
  \& Okamoto}{2006}]{no06} Nagashima M., Okamoto T., 2006, ApJ, 643, 863 

\bibitem[\protect\citeauthoryear{{Navarro} \& {Steinmetz}}{{Navarro} \&
  {Steinmetz}}{1997}]{ns97}
{Navarro} J.~F.,  {Steinmetz} M.,  1997, ApJ, 478, 13

\bibitem[\protect\citeauthoryear{Okamoto 
  \& Nagashima}{2003}]{on03} Okamoto T., Nagashima M., 2003, ApJ, 587, 500 

\bibitem[\protect\citeauthoryear{{Quinn}, {Katz} \& {Efstathiou}}{{Quinn}
  et~al.}{1996}]{qke96}
{Quinn} T.,  {Katz} N.,    {Efstathiou} G.,  1996, MNRAS, 278, L49

\bibitem[\protect\citeauthoryear{Rasera 
  \& Teyssier}{2006}]{rt06} Rasera Y., Teyssier R., 2006, A\&A, 445, 1 

\bibitem[\protect\citeauthoryear{{Rees}}{{Rees}}{1986}]{rees86}
{Rees} M.~J.,  1986, MNRAS, 218, 25P

\bibitem[\protect\citeauthoryear{{Shapiro}, {Giroux} \& {Babul}}{{Shapiro}
  et~al.}{1994}]{sgb94}
{Shapiro} P.~R.,  {Giroux} M.~L.,    {Babul} A.,  1994, ApJ, 427, 25

\bibitem[\protect\citeauthoryear{Schaye et al.}{1999}]{sch99} 
Schaye J., Theuns T., Leonard A., Efstathiou G., 1999, MNRAS, 310, 57 

\bibitem[\protect\citeauthoryear{{Somerville}}{{Somerville}}{2002}]{som02}
{Somerville} R.~S.,  2002, ApJ, 572, L23

\bibitem[\protect\citeauthoryear{{Springel}}{{Springel}}{2005}]{gadget2}
{Springel} V.,  2005, MNRAS, 364, 1105

\bibitem[\protect\citeauthoryear{{Springel} \& {Hernquist}}{{Springel} \&
  {Hernquist}}{2002}]{sh02}
{Springel} V.,  {Hernquist} L.,  2002, MNRAS, 333, 649

\bibitem[\protect\citeauthoryear{{Springel}, {Yoshida} \& {White}}{{Springel}
  et~al.}{2001}]{gadget1}
{Springel} V.,  {Yoshida} N.,    {White} S.~D.~M.,  2001, NewA, 6, 79

\bibitem[\protect\citeauthoryear{{Steinmetz} \& {White}}{{Steinmetz} \&
  {White}}{1997}]{sw97}
{Steinmetz} M.,  {White} S.~D.~M.,  1997, MNRAS, 288, 545

\bibitem[\protect\citeauthoryear{{Susa} \& {Umemura}}{{Susa} \&
  {Umemura}}{2004a}]{su04a}
{Susa} H.,  {Umemura} M.,  2004a, ApJ, 600, 1

\bibitem[\protect\citeauthoryear{{Susa} \& {Umemura}}{{Susa} \&
  {Umemura}}{2004b}]{su04b}
{Susa} H.,  {Umemura} M.,  2004b, ApJ, 610, L5

\bibitem[\protect\citeauthoryear{Theuns et al.}{1998}]{the98} 
Theuns T., Leonard A., Efstathiou G., Pearce F.~R., Thomas P.~A., 1998, 
       MNRAS, 301, 478 

\bibitem[\protect\citeauthoryear{{Thoul} \& {Weinberg}}{{Thoul} \&
  {Weinberg}}{1996}]{tw96}
{Thoul} A.~A.,  {Weinberg} D.~H.,  1996, ApJ, 465, 608

\bibitem[\protect\citeauthoryear{{Weinberg}, {Hernquist} \& {Katz}}{{Weinberg}
  et~al.}{1997}]{whk97}
{Weinberg} D.~H.,  {Hernquist} L.,    {Katz} N.,  1997, ApJ, 477, 8

\end{thebibliography}

\appendix

\section{An improved model for gas accretion onto haloes} \label{MODEL_B}

Once density of accreting gas, $\rho_{\rm acc}$, is specified, 
its temperature can be computed as $T_{\rm acc} = T_{\rm eq}(\rho_{\rm acc})$. 
In model A, we assumed that the density of gas accreting onto a halo is
one-third of the virial gas density, i.e. $\rho_{\rm acc} = 1/3 \rho_{\rm vir}(z)$. 
If the temperature at this density , $T_{\rm eq}(\rho_{\rm vir}/3)$, is higher than the halo's virial temperature, $T_{\rm vir}$, 
the gas cannot accrete. 
If merging haloes have an unusually low baryon fraction, it may be that a larger fraction of baryons is available for accretion.
We attempt to include this idea in our model 'B'~, by writing the density of accreting gas as
\begin{equation}
\rho_{{\rm acc}} = \beta \frac{M_{{\rm acc}}}{M_{{\rm b}}^{\rm max}} \rho_{\rm vir},
\end{equation}
where $\beta$ is a free parameter of the order of unity. A halo with baryon poor progenitors has higher 
$M_{\rm acc}/M_{\rm b}^{\rm max}$, and hence higher $\rho_{{\rm acc}}$. 
The temperature of the accreting gas is now evaluated at $\rho_{{\rm acc}}$, and can be higher than the value 
$\rho_{\rm vir}/3$ of model A.
A value of $\beta = 2/3$ improves the predictions of model B over that of the simpler model A.
With this value of $\beta$, models A and B accrete gas at the same density for a halo with mass $M_{\rm c}$, 
which has a baryon fraction $\langle f_{\rm b}\rangle/2$.

\section{Filtering mass} \label{FILTERINGMASS}

\cite{gh98} found that growth of density fluctuations in the gas is
suppressed for comoving wavenumber $k > k_{\rm F}$, where the critical
wavenumber $k_{\rm F}$ is related to the Jeans wavenumber $k_{\rm J}$ by
\begin{equation}
\frac{1}{k_{\rm F}^2(t)} = \frac{1}{D(t)} \int_{0}^{t} {\rm d}t'
\frac{\ddot{D}(t') + 2H(t')\dot{D}(t')}{k_{\rm J}^2(t')}
\int_{t'}^t \frac{{\rm d}t''}{a^2(t'')}\,.
\end{equation}
The physical Jeans wavenumber $k_{\rm J}$ is defined as
\begin{equation}
k_{\rm J} \equiv \frac{a}{c_{\rm s}} (4 \pi G \langle\rho_{\rm tot}\rangle)^{1/2}.
\label{JEANS}
\end{equation}
Here, $D(t)$ and $H(t)$ are the linear growth factor, and Hubble
constant as functions of cosmic time $t$, respectively, and
the sound speed $c_{\rm s}$ is defined as
\begin{equation}
c_{\rm s} = \left(\frac{5}{3} \frac{k_{\rm B} T_0}{\mu
    m_{\rm p}}\right)^\frac{1}{2}\,.
\end{equation}
Note that \citet{gne00} used the volume-averaged IGM temperature,  $\langle T \rangle_V$, instead of $T_0$.  
He compared the corresponding filtering mass
\begin{equation}
M_{\rm F} = \frac{4 \pi}{3} \langle\rho_{\rm tot}\rangle \left(\frac{2 \pi
    a}{k_{\rm F}} \right)^3
    \label{eq:mf}
\end{equation}
with the characteristic mass measured in cosmological simulations of reionization.
He found a good agreement between the characteristic mass and the filtering
mass down to $z = 4$.
\begin{figure}
\begin{center}
\includegraphics[width=7.5cm]{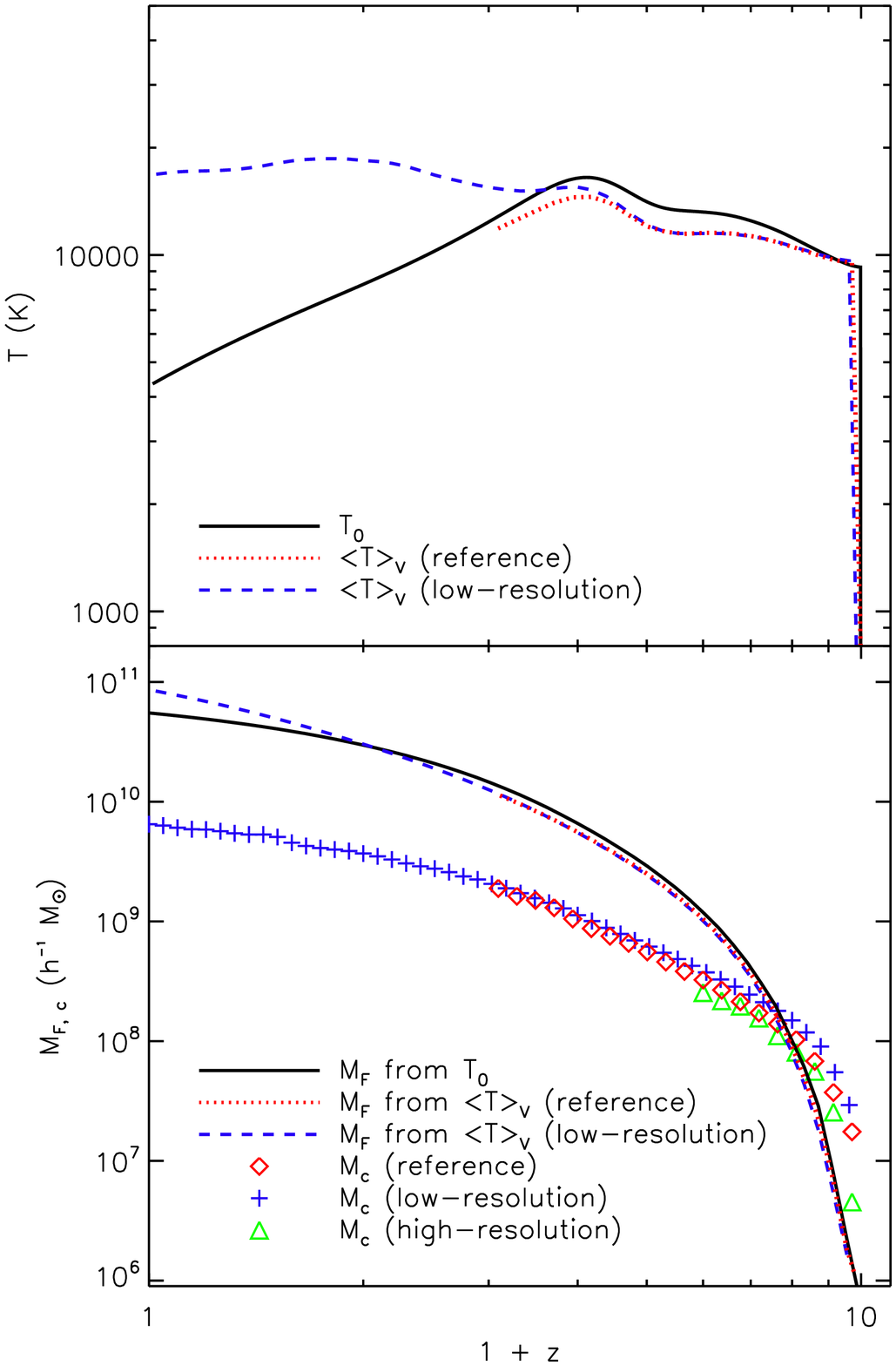}
\end{center}
\caption{
    {\it Upper panel}: Evolution of the IGM temperature at mean density (solid line) and the volume-averaged temperatures in the
    \sm{reference} (dotted line) and \sm{low-resolution} (dashed line) simulations. 
    These temperatures are used to compute the filtering mass shown in the lower panel. 
    {\it Lower panel} : Comparison of
    filtering mass, $M_{\rm F}$, obtained from Eq.~(\ref{eq:mf}), and characteristic mass, $M_{\rm c}$, obtained from the simulations.  The filtering mass $M_{\rm F}\approx 10M_{\rm c}$ by $z=0$, illustrating the fact that the filtering mass strongly overestimates the effect of photoheating on the formation of small galaxies. }
\label{MF}
\end{figure}

In the upper panel of Fig.~\ref{MF}, we show evolution of $T_0$ and the volume-averaged
temperatures in the \sm{reference} and \sm{low-resolution}
simulations. Since larger objects form in the \sm{low-resolution}
simulation owing to its larger box size, the IGM temperature in the
\sm{low-resolution} simulation at low redshift becomes slightly higher
than that in the \sm{reference} simulation because of the shock
heating. In the lower panel, we compare the filtering mass calculated from both $T_0$ and 
$\langle T \rangle_{\rm V}$ 
with the characteristic mass. Clearly, the
filtering mass significantly overestimates the characteristic mass at
low redshift, while there is a reasonably good agreement at high
redshift ($z > 6$) between them.

\bsp
\label{lastpage}
\end{document}